\documentstyle[sprocl]{article}

\def\be{\begin{equation}}
\def\ee{\end{equation}}
\def\bea{\begin{eqnarray}}
\def\eea{\end{eqnarray}}

\begin{document}
\title{THE STRUCTURE OF COSMIC STRING WAKES \cite{wake}}
\author{A. SORNBORGER}
\address{DAMTP, University of Cambridge, 3 Silver Street,
\\ Cambridge CB2 1ND, UK}
\author{R. BRANDENBERGER}
\address{Department of Physics, Brown University,\\
Providence, RI 02912, USA}
\author{B. FRYXELL and K. OLSON}
\address{Institute for Computational Science and Informatics,\\
George Mason University,
Fairfax, VA 22030, USA}
\maketitle\abstracts{We present results of a cosmological
hydrodynamical study of gravitational accretion in cosmic string wakes
and filaments. Cosmic string wakes are formed by fast moving $(v \sim
c)$ strings. A conical deficit angle in the string spacetime induces a
velocity perturbation in the background matter and a two-dimensional
wake accretes in the path of the string. Filaments are formed by
slow moving strings with a large amount of small scale structure. The
major gravitational perturbation from slow moving strings is due to
the Newtonian field induced by the effective mass of the wiggles. In
cosmic string wakes, cool streams of baryons collide and are trapped
at the center of the wake causing an enhancement of baryons versus
dark matter by a factor of $2.4$. In filaments, a high pressure is
induced at the filament core and baryonic matter is expelled leading
to a baryon deficit in the center of the filament.}
\section{Introduction}

Strings are an inevitable result of symmetry breaking in many
theories in particle physics as well as in numerous condensed matter
systems. Once the temperature decreases below the symmetry breaking
scale, the field undergoes a phase transition and a string network
forms. In an expanding universe, the string distribution quickly
approaches a scaling solution in which there are (statistically) the
same number of strings per horizon volume at all times. Numerical
simulations show that the number of strings per horizon volume is in
the range $10 - 30$. Potential systematic errors due to coarse
graining in simulations may tend to favor a lower number of strings
per horizon volume.

If symmetry breaking occurs at the GUT scale, the strings are
massive enough to be responsible for large-scale structure in the
universe. In this case, the scaling solution leads to a
Harrison-Zel'dovich spectrum consistent with that observed by
COBE. Calculations of the power spectrum give good agreement with the
observed power spectrum with a bias factor around $2$. Normalization of
the spectrum with COBE gives a value of $G\mu \sim 10^{-6}$, a
dimensionless quantity proportional to the mass per length of the
string $\mu$. This value is within observational bounds, the most
restrictive being that from timing analyses of the millisecond pulsar.

\section{Large-Scale Structure Formation}

Large-scale structure in the cosmic string model is due to
the growth of density perturbations resulting from the gravitational
field of the string network. The gravitational field of a string can
conceptually be broken into two parts. The first part is due to purely
general relativistic effects: the string leads to a conical
spacetime. Geodesics in such a spacetime converge after passing the
string. This means that a moving string will cause a velocity
perturbation as it passes through matter. For strings with little
small-scale structure such as waves and kinks moving along the string,
the velocity perturbation from the conical spacetime is the
predominant perturbing mechanism. Large-scale structure formed by fast
moving strings will thus be in large sheet-like objects called wakes.

Strings with a large amount of small-scale structure move
slowly. The predominant perturbation mechanism for these strings is
the Newtonian field resulting from the effective mass of the waves and
kinks on the string. These strings move slowly accreting long
cylindrically shaped overdensities called filaments.

\section{Non-linear Matter Evolution (Our Study)}

What has been missing in the study of large-scale structure in the
context of the cosmic string model is an understanding of non-linear
evolution of matter accretion in wakes and filaments. In particular,
what is the detailed distribution of the baryonic and dark matter?

To answer this question, we have developed a high-resolution PPM/PIC
cosmological hydrodynamical numerical code for investigating the
formation and evolution of baryonic and dark matter as they accrete in
cosmic string wakes and filaments.

\section{Results}

We have investigated the matter distribution in the case of wakes and
filaments. The matter distribution in the wake is markedly dissimilar
to that of the filament.

{}First, we investigated the matter distribution in a wake caused by a
long straight string. As the string passes through matter, its conical
spacetime gives a velocity kick to the matter on either side of its
path and a wake forms. Two cool coherent streams of matter flow toward
the center of the wake. The baryonic matter in the streams collides at
the center forming an overdensity bounded by a shock. The dark matter
streams flow through each other giving rise to an overdensity bounded
by caustics. The baryonic matter is more concentrated at the center of
the wake than the dark matter leading to an enhancement of baryonic
matter to dark matter of about $2.4$. Temperatures at the wake center
are of the order of $100$ degrees Kelvin.

Next, we examined the evolution of a filament caused by a slowly
moving string. Here, the matter distribution was quite different than
that of the wake. In the filament, the kinetic energy of the inflow is
higher than that of the wake leading to high post-thermalization
temperatures at the filament core. The associated high pressures force
the baryons out of the filament core resulting in a net baryon deficit
inside the filament. Temperatures at the filament core are of the
order a few times $10^6$ degrees Kelvin.

\section{Conclusions}

{}From our results we conclude that biasing in the cosmic string model
is perturbation dependent. In wakes, biasing of a factor of $2$ to $3$
can be expected, while in filaments, antibiasing is expected. The
amount of biasing in wakes is consistent with calculations of the
amount of biasing required by power spectrum calculations.

We find wakes to be relatively cool, with temperatures of the order
$100$ degrees Kelvin. Filaments are hotter, with temperatures of the
order $10^6$ degrees Kelvin.

The baryon enhancement that we find in wakes is a possible mechanism
for explaining baryon overabundances in objects such as the Coma
cluster. It has been suggested that clusters will form at triple wake
intersections. If this is the case, the enhancement will be tripled to
a factor of $7$, giving a measured baryon fraction of 0.35.

\end{document}